\documentstyle{sup}
\input psfig
\newcommand{\p}{{\bf u}}

\newcommand{\be}{\begin{eqnarray}}
\newcommand{\ee}{\end{eqnarray}}
\newcommand{\bup}{b^{+}}

\newcommand{\sx}{\sigma_x}

\newcommand{\sz}{\sigma_z}

\newcommand{\szzz}{\sigma_3}

\newcommand{\syy}{\tau_y}

\begin{document}
\title[{\it Superlattices and Microstructures, Vol. ??, No. ?, 1999}]{Mesoscopic superconductors under irradiation: Microwave spectroscopy of Andreev states}
\author[{\it Superlattices and Microstructures, Vol. ??, No. ?, 1999}]
{N.~I.~Lundin,
L.~Y.~Gorelik,$^\ast$ R.~I.~Shekhter, and M. Jonson \cr 
{\normalsize\it Department of Applied Physics}\cr
{\normalsize\it Chalmers University of Technology and G{\"o}teborg University,
SE-412 96 G{\"o}teborg, Sweden}\cr
{\normalsize\it $^\ast$and 
B. Verkin Institute for Low Temperature Physics and 
Engineering, 310164 Kharkov, Ukraine}
\vspace{10pt}\cr
V.~S.~Shumeiko 
\cr
{\normalsize\it  Department of Microelectronics and Nanoscience,}\cr
{\normalsize\it Chalmers University of Technology and
G{\"o}teborg University, SE-412 96 G{\"o}teborg, Sweden}
%\cr
%{\normalsize\it and}
%\cr
%{\normalsize\it $^{3}$ Verkin Institute for Low Temperature Physics and
%Engineering,} \cr
%{\normalsize\it National Academy of Sciences of Ukraine, 310164 Kharkov,
%Ukraine} %\vspace{10pt}\cr
% 
}
%\cr
%{\normalsize\it $^{4}$Department of Microelectronics and Nanoscience, Chalmers University of Technology}
%\cr
%{\normalsize\it and G{\"o}teborg University, SE-412 96 G{\"o}teborg, Sweden}\cr}
%{ N.~I.~Lundin, R.~I.~Shekhter, and M.~Jonson%
%\cr
%{\normalsize\it  Department of Applied Physics, Chalmers University of Technology and G{\"o}teborg University,}
%\cr 
%{\normalsize\it SE-412 96 G{\"o}teborg, Sweden}
%\vspace{10pt}\cr
%L.~Y.~Gorelik 
%\cr
%{\normalsize\it  Department of Applied Physics, Chalmers University of Technology and G{\"o}teborg University,}
%\cr
%{\normalsize\it SE-412 96 G{\"o}teborg, Sweden}
%\cr
%{\normalsize\it and}
%\cr
%{\normalsize\it   Verkin Institute for Low Temperature Physics and Engineering,}
%\cr
%{\normalsize\it  National Academy of Sciences of Ukraine, 310164 Kharkov, Ukraine}
%\vspace{10pt}\cr
%
%V.~S.~Shumeiko 
%\cr
%{\normalsize\it Department of Microelectronics and Nanoscience, Chalmers University of Technology}
%\cr
%{\normalsize\it and G{\"o}teborg University, SE-412 96 G{\"o}teborg, Sweden}\cr}

\maketitle
\vspace*{1cm}
\begin{abstract}
We show that irradiation of a voltage-biased
superconducting quantum point contact at frequencies of the
order of the gap energy can 
remove the suppression of subgap dc transport through Andreev levels. 
Quantum interference among resonant scattering events involving
photon absorption is furthermore shown to make
microwave spectroscopy of the Andreev levels feasible. 
We also discuss how the same interference effect can be applied for detecting
weak electromagnetic signals up to the gap frequency, and how it is affected
by dephasing and relaxation.
%
%The dynamics of a superconducting quantum point contact biased at subgap 
%voltages is shown to be strongly affected by a microwave electromagnetic 
%field.
%Interference among a sequence of temporally localized, microwave-induced 
%Landau-Zener transitions between current carrying Andreev levels results 
%in energy absorption and in an increase of the subgap current by several 
%orders of magnitude. The contact is an interferometer in the sense that 
%the current is an oscillatory function of the inverse bias voltage. 
%Possible applications to Andreev-level spectroscopy and microwave 
%detection are discussed. 
%}}
\end{abstract}
%}}}
\vspace*{.5cm}
%\newpage
%\vspace{-1cm}
%\narrowtext

\section{Introduction}
%At this point in time 
It is well known that the current through a voltage-biased
superconducting quantum point contact (SQPC) is carried by localized states. These
states, called Andreev states, 
%bound 
%states, 
%exist in pairs, one above and one under the 
%Fermi level, and 
are confined to the normal region of the
contact. The energy of the states --- the Andreev levels ---
exist in pairs, (one above and one under the
Fermi level), and lie within the
energy gap of the superconductor, with positions which depend on the change $\phi$
in the phase of the superconductors across the junction. The applied
bias affects this phase difference through the Josephson relation, 
$\dot{\phi}=2eV/\hbar$. With a constant applied bias $V$ much smaller than the
gap energy $\Delta$, $\phi$ will increase 
linearly in time, and the Andreev levels will move adiabatically within the gap.
This motion is a periodic oscillation in $\phi$, indicating that no energy is
transfered to the SQPC and a pure ac current will flow through the contact. This
is actually the ac Josephson effect.

We wish to study this system in a non-equilibrium situation, one way to  
accomplish this is by introducing microwave radiation with a frequency $\omega\approx \Delta$,
which will couple the Andreev levels to each other. The radiation will represent
a non-adiabatic perturbation of the SQPC system. However, if the amplitude 
of the electromagnetic field is sufficiently small, the field will not affect
the adiabatic dynamics of the system much unless the condition for resonant
optical interlevel transitions is fulfilled. Such resonances will only occur 
at certain moments determined by the time evolution of 
the Andreev level spacing. The resonances will provide a mechanism 
for energy transfer to 
the system to be nonzero when averaged over time and hence for 
a finite dc current through the junction. The rate of energy transfer 
is in an essential way
determined by the interference between different scattering events 
\cite{ourPRL,us}, and therefore it is not surprising that oscillatory 
features appear in the (dc) current-voltage characteristics of an irradiated SQPC. 
Dephasing and relaxation will affect the interference pattern and  
may even conspire to produce a dc current flowing in the reverse direction with
respect to the applied voltage bias. The mechanism behind 
this negative resistance 
is very similar to the one responsible for the ``somersault effect'' discussed 
by Gorelik et al. \cite{smslt}.

A peculiar feature of the Andreev bound states, in comparison with 
normal 
bound states, is that they can carry current. This is why microwave-induced
transitions between Andreev levels can be detected by means of transport 
measurements. In fact, it is possible to do Andreev energy-level spectroscopy 
in the sense that the Andreev level spectrum at least in principle can be
reconstructed from a measurement of the microwave-induced subgap current.
Such microwave spectroscopy of Andreev states in mesoscopic superconductors is
the topic addressed in this work.

%
%Although an example of the more general problem of energy level spectroscopy,
%the spectroscopy of Andreev levels has an important specific aspect. Due to
%their ability to carry electric current, detection
%of optical transitions between Andreev levels is possible by means of transport
%measurements. The appearance of a subgap current 
%under resonant radiation can furthermore be used as a sensitive microwave 
%detector.
%

\begin{figure}
\centerline{\psfig{figure=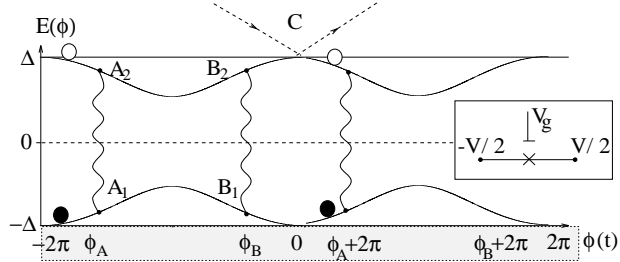,width=8 cm}}
%\vspace{3mm}
 \caption{ \label{fig:junction}
Time evolution of Andreev levels (full lines) in the energy gap of a 
gated voltage-biased, single-mode SQPC (see inset). 
A weak microwave field induces 
resonant transitions (wavy lines) between the levels at points $A$ and $B$ 
and the level above the Fermi energy becomes partly occupied  to an extent 
determined by interference between the two transition amplitudes. 
Non-adiabatic interactions release the energy of quasiparticles in the 
(partly) occupied Andreev level into the continuum 
at point $C$, where the Andreev states and the continuum merge 
into each other (represented by dashed arrows, see text) and the 
initial conditions for the 
Andreev level populations are reset (filled and empty circles). 
}
%\vspace{5mm}
\end{figure}

\section{Theoretical framework}
For an unbiased, mesoscopic SQPC --- which by construction has
a normal region $L$ which is much shorter than the coherence length $\xi_0$
--- the Andreev spectrum of each transport mode has the form  
\begin{equation}
E^\pm(\phi)=\pm E(\phi)=\pm\Delta[1-D\sin^2(\phi/2)]^{1/2}.
\end{equation} 
Here $D$ is the transparency of the mode and the energy is measured from the 
Fermi energy \cite{Fur,Bee}. With a small bias voltage
applied, the levels  move along the adiabatic trajectories 
$E^\pm(t)=\pm E(\phi_0+2eVt/\hbar)$
in energy-time space, oscillating with a period of $T_p=\hbar\pi/eV$,  
as shown in Fig.~\ref{fig:junction}. In equilibrium the lower Andreev level
will obviously be occupied, while the upper level will be empty.
In the discussion below we consider a single-mode
SQPC, although we note that the theory also
applies for the case of a single dominant mode in a multi-mode junction.
The transparency $D$ of the mode is taken to be arbitrary
but energy-independent, $0<D<1$.

A high frequency electromagnetic
field is applied to the gate situated near the contact,
see inset in Fig.~\ref{fig:junction}.
The time-dependent electric field induced by the gate is concentrated 
within the non-superconducting region of the junction, and hence the 
charge carriers will couple to the electromagnetic field only there.
When the criterion $\hbar\dot\phi=2eV\ll 2E^2(t)/\Delta$ 
for adiabaticity is obeyed,
the rate of interlevel transitions is exponentially small keeping the
level populations constant in time~\cite{Ave1,Lv1}. 
The presence of a weak electromagnetic field [on the scale of $ E(t)$]
does not affect the adiabatic level trajectories except for short times
close to the resonances at $t=t_{A,B}$, when $E(t_{A,B})=\hbar\omega/2$.
Here the dynamics of the system is strongly
non-adiabatic with a resonant coupling which effectively mixes
the adiabatic levels. This is an analog of the well known Landau-Zener
transition, which describes interlevel scattering as a resonance point 
is passed. In our case these transitions give rise to a splitting of 
the quasiparticle trajectory at the points $A, B$ into two paths; $A_1A_2B_2$ 
and $A_1B_1B_2$, forming a loop in $(E,t)$ space (see Fig.~\ref{fig:junction}).

The resonant scattering opens a channel for energy absorption by the system;
a populated upper 
%Andreev 
level when approaching the edge of the 
energy gap (at point $C$ in Fig.~\ref{fig:junction}) creates real excitations 
in the continuum spectrum, which carry away
the accumulated energy from the contact. As a result, the net rate of 
energy transfer to the system is finite; it consists of energy absorbed 
both from the electromagnetic field and from the voltage source. 
The confluence of the two adiabatic trajectories at $B_2$ 
gives rise to a strong interference pattern in the probability for 
real excitations at the band edge, point $C$. The interference effect is 
controlled by the difference of the phases acquired by the system during 
propagation along the paths $A_1A_2B_2$ and $A_1B_1B_2$.

\subsection{Bogoliubov-de Gennes equation}
In order to describe the time evolution of the Andreev states in more detail
we use the time-dependent Bogoliubov-de Gennes, (BdG), equation \cite{book}
for the quasiclassical envelopes $u_\pm(x,t)$ of the two-component wave function
$\Psi(x,t)=u_+(x,t)e^{i k_Fx} + u_-(x,t)e^{-i k_Fx}$,
\begin{equation}
i\hbar \partial{\bf u}/\partial t =[{\bf H}_0 + \sigma_z V_g(x,t)] {\bf u} \, .
\label{BdG}
\end{equation}
In this equation ${\bf u}=(u_+, u_-)$ is a four-component vector, while
${\bf H}_0$ is the Hamiltonian of the electrons in the electrodes of the point
contact,
\begin{equation}
%\begin{eqnarray}
\label{H0}
{\bf H}_0
%&=&
=
-i\hbar v_F\sigma_z\tau_z\partial/\partial x + 
%\\
%&&
\Delta[\cos(\phi(t)/2)\sigma_x + i\sin(\phi(t)/2)\mbox {sgn} x\sigma_y],
%\nonumber
%\end{eqnarray}
\end{equation}
where $\sigma_i$ and $\tau_i$ denote Pauli matrices in
electron-hole space and in $u_\pm$ space respectively.
For a mesoscopic junction the gap function $\Delta({\bf r})$ need
not be calculated self-consistently, which is why in Eq.~(\ref{H0}) 
it is assumed
to have the constant value $\Delta\exp(i\phi/2)$ in the superconducting
reservoir on one side of the SQPC and the different constant value 
$\Delta\exp(-i\phi/2)$ in the reservoir on the other side.
The gate potential $V_g(x,t)=V_\omega(x)\cos \omega t$
in Eq.~(\ref{BdG}) oscillates rapidly in time and
the amplitude 
%of the oscillations 
is assumed to be
small compared to the Andreev level spacing, $V_\omega\ll E(t)$.

The function ${\bf u}$, which is smooth on the scale of the Fermi wavelength, 
has a discontinuity at $x=0$, i.e. at the point contact, whose spatial 
extension can be neglected. The discontinuity translates to a boundary
condition for ${\bf u}$, 
\begin{equation}
{\bf u}(+0)=(1/\sqrt D)(1-\sqrt R\tau_y){\bf u}(-0), \;\; R=1-D,
\label{BC}
\end{equation}
which can be found \cite{Sh} by matching at the point $x=0$ 
two scattering state solutions to the BdG equation approaching 
from left and right.
%
%which is determined by the transfer
%matrix of the QPC in the normal state and is described by the following 
%boundary condition~\cite{Sh}. It is found by matching two solutions to 
%the BdG equation through the scattering region at $x=0$:

\subsection{Resonant coupling of Andreev levels}
Under the assumption that $eV,V_g\ll \Delta$ the system of Andreev levels
experiences an 
adiabatic evolution in time. This is true at all times except close to 
the resonances (points $A$ and $B$ in Fig.~\ref{fig:junction}) and at 
point C, which will be discussed later.
%The duration $\delta t$ of these resonances is short in the limit 
%$eV\ll\Delta$ compared to the period of Josephson oscillations
%$T_p=\pi\hbar/eV$.
For a resonant transition to occur, the deviation 
of the interlevel spacing from the resonance value
$E(t)-\hbar\omega/2=\dot E(t_{A,B})\delta t$, where $\delta t$ is the 
duration of the resonance,  has to be less than the
quantum mechanical resolution of the energy levels $\hbar/\delta t$. Using
this criterion
we can estimate the duration $\delta t$ of the non-adiabatic evolution as
$\delta t\sim [\hbar/\dot E(t_{A,B})]^{1/2}$, which implies that
$\delta t$ is much shorter than 
the period $T_p=\hbar\pi/eV$ of Josephson oscillations if $eV\ll\Delta$. 
Hence, if this inequality is obeyed we may
consider the non-adiabatic dynamics as temporally localized scattering events.
We shall use this later to derive an analytical result for the induced dc
current. For the time being, however, we will keep the discussion a little more
general. We therefore introduce 
an Ansatz for the wave function {\bf u} in the normal region of the junction
in terms of a linear 
combination of the eigenstates ${\bf u}^\pm$ corresponding to the adiabatic 
Andreev levels $E^\pm$. The rapidly oscillating terms are explicitly 
introduced in this so called resonance approximation, 
\begin{equation}
{\bf u}(t)=b^+ (t)e^{-i\omega t/2}{\bf u}^+ + 
b^-(t) e^{i\omega t/2}{\bf u}^- \, .
\label{b}
\end{equation}
Following Ref~\cite{smslt} we insert the Ansatz (\ref{b}) into the BdG 
equation (\ref{BdG}) to find, using the notation $\vec{b}=(b^+,b^-)$, 
an equation for the coefficients $b^\pm$,
\begin{equation}
i \dot{\vec{b}}(t)=
\left[ \begin{array}{cc} \delta \omega_{+-}(t) & V_{+-}/\hbar
 \\ V_{+-}^\dagger/\hbar & -\delta \omega_{+-}(t) \end{array} \right] 
\vec{b}(t).
\label{eq.coupled}
\end{equation}
Here $\delta \omega_{+-}(t)=[2E^+(t)-\hbar\omega]/2\hbar$ is a measure
of the 
deviation from resonance. This equation describes the time evolution of 
the coefficients $b^+$ and $b^-$, which embody the dynamics of the 
population of the Andreev states under irradiation. 
We recall that the coupling to the time-dependent field is finite only 
in the normal region, which explains why $V_{+-}$ in (\ref{eq.coupled}) 
obtains as the matrix element of $\sigma_z V_g(x,t)$ between the Andreev
states ${\bf u}^\pm$; $V_{+-}=(V_\omega/2)\int dx \langle\p^+\sz\p^-\rangle$,
where $\langle..\rangle$ denotes a scalar product in 4-dimensional space.
For the case of a double barrier SQPC structure $V_{+-}$ was calculated 
in Ref.~\cite{smslt}. For a single barrier
junction an analogous calculation gives us, 
$V_{+-}=\alpha(L/\xi_0)\sqrt{DR}V_\omega\sin(\phi/2)$, where
the constant $\alpha\sim 1$ is determined by the position of the barrier.
We note, that this matrix element 
is proportional to the {\em reflectivity} of the junction; reflection mixes
electron states with $+k_F$ and $-k_F$ allowing optical transitions between
the Andreev levels. In a perfectly transparent SQPC ($D=1$), the upper and
lower Andreev levels correspond to opposite electron momenta 
--- the two levels cannot be coupled by radiation since momentum cannot be
conserved --- 
and the effect under consideration does not exist, cf. Refs. \cite{Ave1,Zai}.

\subsection{Boundary conditions at $\phi=2\pi n$}
Before we proceed to calculate the current through the irradiated SQPC
we need to discuss the boundary condition at $\phi=2\pi n$ ($n$ is an integer). 
In the vicinity of
these points, the Andreev levels approach the continuum and the adiabatic
 approximation is
unsatisfactory, even at small applied voltages and weak electromagnetic fields.
The duration $\delta t$ of the non-adiabatic interaction between the Andreev
level and the continuum states can be estimated 
using the same argument as for the microwave-induced Landau-Zener scattering. 
One finds that, $\delta t \sim \hbar/(\Delta e^2V^2)^{1/3}$. If we evaluate 
the condition, $\delta t<<T_p$, we find that we would require that $eV/\Delta\ll
\pi^3$. This means that we can safely treat the non-adiabatic region as short.
To derive the boundary condition, for example, at point $C$ in
Fig.~\ref{fig:junction}, one needs to calculate the transition amplitude
connecting the states  ${\bf u^+}(t_1)$ at time  $t_1\ll t_C-\delta t$ and ${\bf
u^+}(t_2)$ at time $t_2\gg t_C+\delta t$:  $\langle {\bf u^+}(t_2)\;{\bf
U}(t_2,t_1){\bf u^+}(t_1)\rangle$. Here ${\bf U}(t_2,t_1)$ is the exact
propagator corresponding to the Hamiltonian  in Eq. (\ref{BdG}). We will now
proceed with symmetry arguments to show that  this amplitude is exactly zero.

It can be shown that both the Hamiltonian (\ref{H0}) and the 
boundary condition for ${\bf u}$ at $x=0$ (\ref{BC})
are invariant under the simultaneous charge- and parity inversion
described by the unitary  operator ${\bf\Lambda}\equiv\hat P\sigma_x\tau_z$,
where $\hat P$ is the parity operator in $x$-space. This implies that 
at any time any non-degenerate eigenstate of the 
Hamiltonian is an eigenstate of 
the symmetry operator ${\bf\Lambda}$ with eigenvalue $+1$ or $-1$ and that
this property persists during the time evolution of the state. Specifically,
if we take a state on each side of $\phi=0$,
\begin{eqnarray}
{\bf \Lambda}{\bf u}^+(\phi)&=&\lambda_1 {\bf u}^+(\phi)\label{sym1}\\
{\bf \Lambda}{\bf u}^+(-\phi)&=&\lambda_2 {\bf u}^+(-\phi)\label{sym2}.
\end{eqnarray}
Next we insert ${\bf u}^+(-\phi)=\Upsilon {\bf u}^+(\phi)$, where the 
operator $\Upsilon=\sx\syy$ changes the sign of the phase $\phi$, into 
Eq.~(\ref{sym2}) and apply $\sx\syy$ from the left and arrive at,
\begin{eqnarray}
{\bf \Lambda}{\bf u}^+(\phi)=-\lambda_2 {\bf u}^+(\phi),
\end{eqnarray}
which shows that $\lambda_1=-\lambda_2$. This means that the two 
%eigenstates,
states, 
${\bf u}^+(\phi)$ and ${\bf u}^+(-\phi)$ are orthogonal.
% for all $t$. 
%Especially at $t\gg \delta t$, and we can conclude that 
%$\langle{\bf u}^+(t_2)U(t_2,t_1){\bf u}^+(t_1)\rangle=0$. We have shown that 
%the two states are orthogonal even though they belong to the {\it same} 
%energy level. 
This is consistent with the results of Shumeiko et.al.~\cite{Sh},
 who have shown that the Andreev state wave functions are 
$4\pi$-periodic whereas the energy levels and the current are
$2\pi$-periodic. 
%
%The orthogonality property shown above guarantees that the population of 
%our system has a period consisting of one Josephson oscillation and that 
%the equilibrium population of the Andreev levels is
%reset at each point $\phi=2\pi n^+$, in agreement with Averin and 
%Bardas~\cite{Ave}. This imposes the boundary
%condition $b^+(2\pi n+0)=0,\;b^-(2\pi n+0)=1$ in the beginning of each
%period at $t=0$ or $\phi=0$.
%
%In particular,
%the static Andreev state obeys the equation 
%${\bf \Lambda}{\bf u^+}(\phi)=\Lambda{\bf u^+}(\phi)$
%at any $\phi$. It follows from Eq.~(\ref{H0}) that 
%${\bf u^+}(-\phi)=\sigma_x\tau_y {\bf u^+}(\phi)$ so
%$\Lambda(-\phi)=-\Lambda(\phi)$. The immediate consequence of this property 
%is that the static Andreev states correspond to {\em different}
%eigenvalues $\Lambda$ at $\phi<0$ and at $\phi>0$, and therefore they are 
%orthogonal \cite{period}, even though they belong to the {\em same}
%level.
%
Since the state evolving from the adiabatic state ${\bf u^+}(t_1)$ is
orthogonal to the adiabatic state ${\bf u^+}(t_2)$,
%* because
%of the fact that they are eingenfunctions of operator ${\bf \Lambda} with different eigenvalues. 
%a result
the probability
for an adiabatic Andreev state to be ``scattered" into a
localized state after passing the non-adiabatic region is identically zero.
In reality,
the Andreev state as it approaches the continuum band edge 
decays into the states of the continuum.  Such a decay corresponds to a
delocalization in real space and is the mechanism for
transferring energy to the reservoir \cite{LV}.

The orthogonality property shown above guarantees that the coherent
evolution of our system persists during only one period of the Josephson
oscillation and that the equilibrium population of the Andreev levels is
reset at each point $\phi=2\pi n$ \cite{Ave}. This imposes the boundary
condition
\begin{eqnarray}
b^+(2\pi n+0)=0,\;b^-(2\pi n+0)=1,
\label{bcphi}
\end{eqnarray}
at the beginning of each period. 

\subsection{Calculation of the current}
The quasiclassical equation for the total time dependent current at the
junction ($x=0$) reads
$I(t)=-e v_F\langle{\bf u}\tau_z{\bf u}\rangle\mid_{x=0}$.
%
%An equivalent form for the current can be found from 
%Eqs.~(\ref{BdG},\ref{H0}),
%
By manipulating the BdG equation as given by (\ref{BdG}) and (\ref{H0}) 
and omitting 
%small contributions from rapidly 
oscillating terms which will not contribute to the dc current --- 
the quantity of interest --- one may
turn this expression into the following form,
\begin{equation}
I(t)=\frac{2e}{\hbar}\left({d\phi\over dt}\right)^{-1}\int^\infty_{-\infty}
dx\left[ i\hbar \frac{d\langle{\bf u}\dot {\bf u}\rangle}{dt}-
\dot V_g\langle{\bf u}\sigma_z{\bf u}\rangle
\right].
\label{j}
\end{equation}
In the static limit, $\dot V_g=0$ and $\dot\phi\rightarrow 0$, this result
is clearly equivalent to the standard 
expression $I=(2e/\hbar)\,(dE^\pm/d \phi)$
for the Andreev level current. 

In the general nonstationary case ${\bf u}$ 
is a linear combination of ${\bf u}^\pm$ according to Eq.~(\ref{b}) and we
%
%, we use
%the linear combination for to 
%
can calculate the current averaged over the period $T_p$.
This is done through Eq.~(\ref{j}) with help of Eq.(~\ref{eq.coupled}) and 
the normalization condition $|b^+|^2+|b^-|^2=1$, leading to
\begin{equation}
I_{dc}={2e\over \pi\hbar}\left(\Delta-{\hbar\omega\over 2}\right)\left[
|b^+(T_p)|^2-|b^+_0|^2\right],
\label{Idc}
\end{equation}
where $|b^+(T_p)|^2$ is the population of the upper level at the 
end of the period and $|b^+_0|^2$ at the beginning of the period. The
result simplifies further since 
according to Eq.~(\ref{bcphi}) $|b^+_0|^2=0$.

The direct current through the contact can be viewed as resulting from
photon-assisted pair tunneling or equivalently as being due to the
distortion of the ac pair current due to the induced interlevel
transitions.
%The magnitude of the current is such that the energy absorbed from
%the voltage source, $V\cdot I_{dc}$, together with the energy absorbed
%from the hf field corresponds to the energy necessary for creating a
%real continuum-state excitation.
%
To facilitate an understanding of the current expression in another way, 
let us study 
energy conservation. We have two sources of energy, the applied field and the
applied bias. If we consider a single Josephson period $T_p$, the energy
absorbed by the system from the voltage source is $I\cdot V\cdot T_p$, and from
the applied field it is $\hbar \omega |\bup(T_p)|^2$ (assuming for simplicity
that $|b^+_0|^2=0$). Energy conservation for the period can be stated as,
$E_{out}=E_{absorbed}$, with the energy leaving the system stated as $2 \Delta
|\bup(T_p)|^2$: 
\begin{eqnarray} 
2\Delta |\bup(T_p)|^2=I\cdot V\cdot T_p + \hbar \omega
|\bup(T_p)|^2. 
\end{eqnarray} 
The current can then easily be found as, 
\begin{eqnarray} 
I= \frac{1}{V
T_p}(2\Delta |\bup(T_p)|^2-\hbar \omega |\bup(T_p)|^2) \end{eqnarray} or, inserting the
value of $T_p$, 
\begin{eqnarray}
I=\frac{2e}{\hbar \pi}(\Delta -\frac{\hbar \omega}{2}) |\bup(T_p)|^2.
\end{eqnarray}
This is exactly the same result as obtained by the more tedious method used above. This discussion allows us to interpret the current 
as voltage-bias mediated, photon assisted tunneling.

\section{Results of numerical and analytical analysis}
Generally the current is given by Eq.~(\ref{Idc}) with the boundary 
condition Eq.~(\ref{bcphi}) inserted. The actual calculation of the current 
has to be done numerically in most cases. However, in the limit of a weak 
applied external field and when the frequency of the applied field is such 
that we have two (in time) well separated resonances, we can treat the 
resonances as temporally localized scattering events. This approach
can be quite rewarding as will become clear below. 

\subsection{Scattering formalism}
Formally, we can describe the system's evolution through a resonance by letting
a scattering matrix $\hat S$ connect the
coefficients $b^\pm$ before and after the 
splitting points $A$ and $B$. Approximating the time dependence of the Andreev
levels to be linear, a standard analysis of the Landau-Zener
interlevel transitions (see e.g. \cite{smslt}) gives the scattering matrix
elements at the points $A$ and $B$ as,
%$(S_A)_{++}=(S_A)_{--}=\tau$,
%$(S_A)_{+-}=-(S_A)^\ast_{-+}=\rho $, 
\begin{eqnarray}
\hat{S}_A=\left[ \begin{array}{cc} r&-d e^{-i \Theta} \\ 
d e^{i \Theta}&  r \end{array} \right],
\hat{S}_B= \left[ \begin{array}{cc}r &  d e^{i \Theta} \\ 
                 -d e^{-i \Theta}&  r 
 \end{array} \right],
\label{sc_mat}
\end{eqnarray}
where $|d|^2 =1-|r|^2=1-e^{-\gamma}$ is the probability of the 
Landau-Zener interlevel transition.
Here $\gamma=\pi |V_{+-}|^2/|dE/dt|$
where $V_{+-}$ is once again the matrix element for the interlevel transitions.
%
%The time-dependent electric field induced by the gate is concentrated
%within the non-superconducting region of the junction, and therefore the
%explicit form of the matrix element $V_{+-}$ depends on the concrete model of 
%this region.  Assuming a ballistic normal region with a single
%impurity scatterer, we directly calculate the matrix element of operator 
%$\sigma_z V_g(x,t)$ between the eigenstates ${\bf u}^\pm$, 
%$V_{+-}=\alpha(L/\xi_0)\sqrt{DR}V_\omega\sin(\phi/2)$, where
%$L$ is the length of the normal region, $\xi_0$ is 
%the coherence length, and the
%constant $\alpha\sim 1$ is determined by the position of the impurity 
%\cite{matrel}. 
%

\begin{figure}
\vspace{-.8cm}
%\vspace{45mm}
\centerline{\psfig{figure=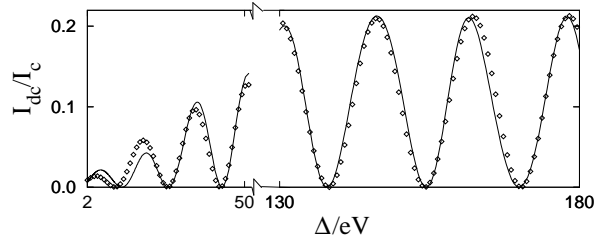,width=8 cm}}
 \caption{ \label{fig:compare}
Current vs. inverse voltage from Eq.~(15) for a biased SQPC 
irradiated with microwaves of frequency $\omega=1.52\Delta/\hbar$ and 
amplitude corresponding to a matrix element $|V_{+-}|=0.024\Delta$ 
for interlevel transitions, $I_c=e\Delta D/\hbar$. 
Note the cut in the inverse voltage scale.
Results of the scattering approach, 
Eq.~(19) [$\diamond$], are close to those obtained by numerically solving 
the BdG-equation with the radiation field treated in the resonance 
approximation (solid line).
The close fit means that the scattering picture can be used to reconstruct
the Andreev levels from the period of the current oscillations (Andreev level 
spectroscopy, see Section 3.2). 
}
%\vspace{5mm}
\end{figure}

By introducing the matrix $\hat\Phi_{j,i}=\exp{(i\sigma_z\Phi(i,j))}$, 
\begin{equation}
\Phi(i,j)= {1\over 2eV}\int_{\phi_i}^{\phi_j}
d\phi \left( E(\phi)-{\hbar\omega\over 2}\right),
\label{phi}
\end{equation}
which describes the ``ballistic" 
dynamics of the system between the Landau-Zener
scattering events, we connect the coefficients $b^\pm$ at the end of the
period of the Josephson oscillation, $\phi=0$, with the coefficients 
$b^\pm_0$ at the beginning of the period, $\phi=-2\pi$,
\begin{equation}
{b^+\choose b^-}=\hat\Phi_{0,B}\hat S_B\hat\Phi_{B,A}\hat S_A\hat\Phi_{A,-2\pi}
{b^+_0\choose b^-_0}.
\label{bb0}
\end{equation}
%
%${{\bf b}}=\hat\Phi_{B2\pi}\hat S_B\Phi_{AB}\hat S_A\Phi_{0A}{{\bf b}}_0$.
The time-averaged current through the junction can be directly expressed 
through these coefficients.
Combining the boundary condition (\ref{bcphi}) with Eqs.~(\ref{Idc}) and
(\ref{bb0}), one finds
\begin{equation}
I_{dc}={8e\over \pi\hbar}\left(\Delta-{\hbar\omega\over 2}\right)
e^{-\gamma}(1-e^{-\gamma})\sin^2(\Phi(A,B)+\theta),
\label{I}
\end{equation}
where $\theta$ is the phase of the probability
amplitude for the Landau-Zener
transition, which only weakly depends on $V$. In Fig.~\ref{fig:compare} the 
current is calculated both numerically and with the expression above, 
Eq.~(\ref{I}). The current is plotted as a function of inverse voltage in 
order to show a periodicity in $1/V$. The close fit in 
Fig.~\ref{fig:compare} ensures us that the scattering approach works 
well for the 
situation when the resonances are well separated and the applied field is weak.

\subsection{The quantum point contact as an interferometer}
Equation (\ref{I}) is the basis for presenting the biased SQPC 
as a quantum interferometer. There is a clear analogy between the 
SQPC interferometer and a standard SQUID in that they both rely on the presence 
of trajectories that form a closed loop. In a SQUID, which is used to measure
magnetic fields, the loop is determined by the device geometry; in the SQPC the 
voltage (analog of the magnetic field) is well defined while the ``geometry"
of the loop in $(E,t)$ space can be measured. This loop is determined by
the Andreev-level trajectories in $(E,t)$ space and is controlled by the
frequency of the external field. This gives us an immediate possibility to
reconstruct the phase dependence of the Andreev levels from the frequency
dependence of the period $\Pi$ of oscillations of the current versus inverse 
voltage, see Fig.~2. Indeed, it follows from Eq.~(\ref{phi}), that
\begin{equation}
\phi(E)=\pi\pm {4\pi e\over\hbar}
\left.{d\Pi^{-1}\over d\omega}\right|_{\omega=2E}.
\end{equation}
This shows that the position of the Andreev energy levels can be reconstructed
from measurements of microwave-induced subgap currents through the SQPC.

\section{Discussion - dephasing and relaxation}
In order to be able to do interferometry it is necessary to keep
phase coherence during at least
one period of the Josephson oscillation. There are
three dephasing mechanisms that impose limitations in practice: (i)
deviations from an ideal voltage bias, (ii) microscopic interactions, and
(iii) radiation induced transitions to continuum states. The main source of
fluctuations of the applied voltage is the ac Josephson effect.
According to the RSJ model, a fixed voltage across the
junction can only be maintained if the ratio between the intrinsic 
resistance $R_i$ of the voltage source and the normal junction
resistance $R_N$ is small. If
$R_i/R_N\ll 1$ the amplitude of the voltage fluctuations $\delta V$ is
estimated as $\delta V\sim (R_i/R_N)\Delta$. Effects of voltage fluctuations
on the accumulated phase $\Phi(A,B)$ can be neglected if 
$\delta \Phi(A,B)=\Phi'(A,B)\delta V\ll 2\pi$, i.e. if $eV>(R_i/R_N)^{1/2}
\Delta$,
which corresponds to a lower limit for the bias voltage \cite{C}.

The dephasing time due to microscopic interactions is comparable to the
corresponding relaxation time \cite{Gefen}. This mechanism of dephasing 
can be neglected as soon as the relaxation time exceeds the Josephson 
oscillation period, $\tau_i\gg \hbar\pi/eV$. 
Taking electron-phonon interaction as the leading mechanism of inelastic
relaxation, we estimate $\tau_i$ to be of the order of the electron-phonon
 mean free time at the critical temperature, $\tau_{ph}(T_c)$, since the
large deviations from equilibrium in our case occur in the energy interval
$E<\Delta$. This gives \cite{Kaplan} another limitation on how small 
the applied voltage can be, $eV > 10^{-2}\Delta$.

The third mechanism of dephasing becomes important when the Andreev 
levels are closer 
than $\hbar \omega$ to the continuum band edge. One can estimate the 
corresponding relaxation time as $\tau_\omega \sim \hbar \Delta/V^2_\omega$. 
For small radiation amplitudes $\tau_\omega$ exceeds $T_V$, while for 
optimal amplitudes they are about equal. The effect of the level-continuum 
transitions on the interference oscillations depends on the frequency.
If $\hbar \omega < 2\Delta /3$ the ``loop region" [$E(t) < \hbar \omega /2$]
is optically disconnected from the continuum, and 
transitions cannot destroy interference. Possible level-continuum transitions
at times outside the loop will only decrease the amplitude of the effect by a
factor $\exp(-\alpha V_\omega^2/eV\Delta)$, where $\alpha < 1$ is the relative 
fraction of the period $T_V$ during which transitions to the
continuum are possible. Accordingly, this factor
is of the order of unity for the voltages that correspond to the
maximum amplitude of oscillations. If $\hbar \omega > 2\Delta /3$,
the interference is impeded by the optical transitions into the
continuum, and the current oscillations decrease. Still,
a nonzero average current through the junction will persist.

In any real system, both relaxation and dephasing will be present as 
discussed above. Therefore, to complete our analysis we will simulate 
these effects on our system. If we assume that the relaxation and dephasing 
times, $\tau, \tau_\phi$ are 
long compared to the duration of the 
non-adiabatic resonances, $\delta t$, we can use a technique~\cite{Gefen}, 
where dissipation is modelled by adding a term to 
the time evolution equation of 
the density matrix for the two-level Andreev system. The equations are, 
where relaxation enters in the diagonal terms and dephasing through the 
off-diagonal terms,  
\begin{eqnarray}
&&\dot \rho_{nn}(t)=-\frac{i}{\hbar}[H_0(t),\rho(t)]_{nn}-
\frac{\rho_{nn}-\rho_{nn}^{eq}}{\tau },\label{eq.d1} \\
&&\dot \rho_{nn'}(t)=-\frac{i}{\hbar}[H_0(t),\rho(t)]_{nn'}-
\frac{\rho_{nn'}}{\tau_{\phi}}, \; \mbox{n$\neq$n'}.
\label{eq.d2}
\end{eqnarray}
with $\tau $ as the characteristic time for relaxation of the system 
to the equilibrium population, $\rho^{eq}$ and $\tau_\phi$ as the 
characteristic time for dephasing. These equations describe the ``ballistic'' 
dynamics of the dissipative system, replacing Eq.~(\ref{phi}).

%The resonant events which are unaffected by the dissipation  can still be 
%modelled with the scattering matrices introduced earlier, Eq.~(\ref{sc_mat}),
%\begin{eqnarray}
%\rho'=\hat{S}_{A,B}\; \rho\; \hat{S}_{A,B}^\dagger.
%\end{eqnarray}

The exact form of the density matrix for the system considered here will be 
a $2{\times}2$ matrix for the discrete two level space of the Andreev states. 
(To avoid confusion we will use $\sigma_1, \sigma_2, \sigma_3$ to denote the 
Pauli matrices in this discrete space.) The Hamiltonian for the density 
matrix is found in Eq.~(\ref{eq.coupled}), and the diagonal elements 
$\rho_{11}$ and $\rho_{22}$ will represent the population of the upper and 
lower Andreev levels.
The resonant events which are unaffected by the dissipation  can still be
modeled with the scattering matrices introduced earlier, Eq.~(\ref{sc_mat}),
\begin{eqnarray}
\rho'=\hat{S}_{A,B}\; \rho\; \hat{S}_{A,B}^\dagger.
\end{eqnarray} 

To calculate the effect of relaxation and dephasing on the current we will 
use the standard expression, $I=2e/\hbar(\partial E/\partial\phi)$, for the 
current carried by a populated Andreev state and average over one Josephson 
period. The result is, as a function of the upper levels population, 
\begin{eqnarray} 
I_{dc}^{diss}=\frac{1}{T_p}\int_0^{T_p} I(t)dt=
\frac{1}{T_p}\frac{2e}{\hbar}\int_0^{T_p} 
Tr(\sigma_3 \rho)\partial E/\partial\phi\; dt=\frac{1}{T_p V}\int_0^{T_p} 
\frac{\partial E^+(t)}{\partial t}\left[2 \rho_{11}-1\right]dt,
\label{eq.dispcurr}
\end{eqnarray}
where we have used the Josephson relation $\dot{\phi}=2eV/\hbar$ and the 
conservation of probability, $Tr (\rho)=1$.

By solving the time evolution equations of the density matrix for the whole 
Josephson period and imposing the boundary condition $\rho(0)=\rho^{eq}=(1-\szzz)/2$ 
%at $t=0$, 
,we arrive at the following result,

\begin{eqnarray}
\begin{raggedleft}
\rho_{11}=
\left\{
\begin{array}{lll}
0& \mbox{, $0< t< t_1$}& \\
d^2\; e^{-(t-t_1)/\tau }& \mbox{, $t_1< t < t_2$}&\\
f(r^2,\tau_\phi,\tau )\;e^{-(t-t_2)/\tau } &\mbox{, $t_2 < t < T_p$}&
\end{array} \right.
\end{raggedleft}
\end{eqnarray}
where
\begin{eqnarray}
f(r^2,\tau_\phi,\tau )=\left[ d^2 (1-e^{-(t_2-t_1)/\tau })+
2r^2 d^2 (e^{-(t_2-t_1)/\tau }-e^{-(t_2-t_1)/\tau_\phi} \cos{2(\Phi(A,B)+
\Theta)})\right].
\end{eqnarray}
Inserting these expressions into Eq.~(\ref{eq.dispcurr}) we find,
\begin{eqnarray}
&&I_{dc}^{diss}(V)=\frac{2}{VT_p}\left[ \frac{d^2\hbar\omega}{2}
\left(e^{-(t_2-t_1)/\tau }-1\right) + \frac{d^2}{\tau }
\int_{t_1}^{t_2}E^+ (t) dt + \right.\nonumber\\
\nonumber&&\left. f(r^2,\tau_\phi,\tau ) \left\{
\left(\Delta e^{-(T_p-t_2)/\tau } -\frac{\hbar\omega}{2}\right)+ 
\frac{1}{\tau } \int_{t_2}^{T_p}E^+(t) e^{-(t-t_2)/\tau } 
dt \right\}\right].
\end{eqnarray}

\begin{figure}
% \begin{center}
   \centerline{\psfig{figure=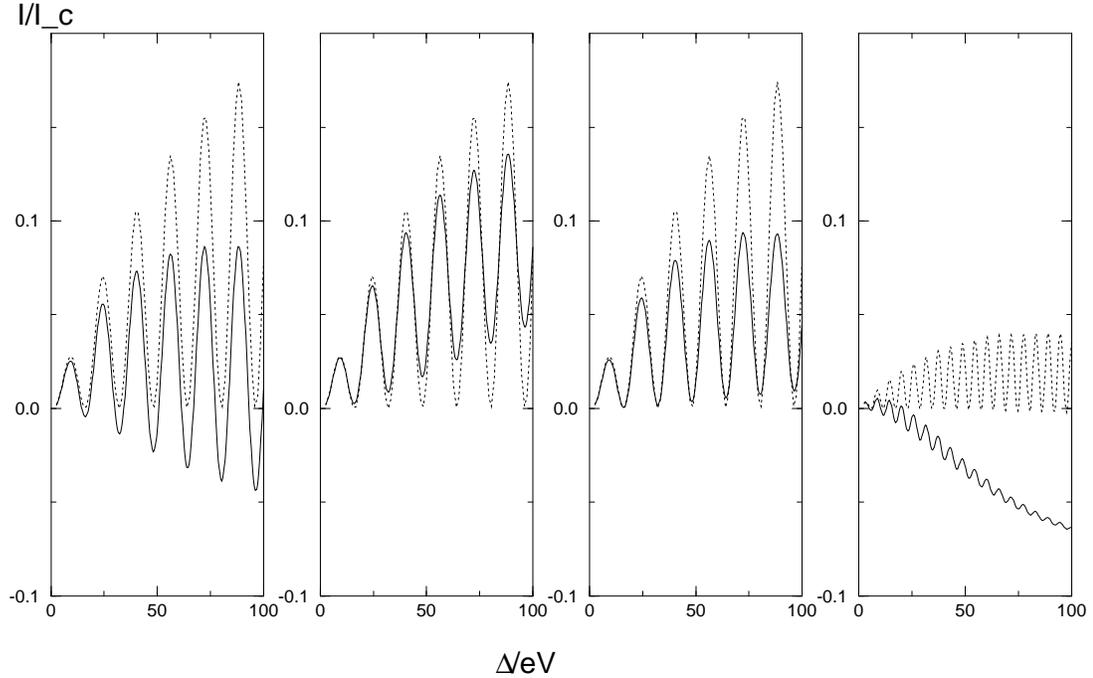,width=14.cm}}
   \caption{Simulations of the current when relaxation and dephasing are
added. In a) only relaxation is added, with $\tau \approx 2 
\cdot 10^{-10}$s, the average value of the current decreases and also may 
become negative. In 
b) there is only dephasing, with $\tau_\phi\approx 5 
\cdot 10^{-10}$s. In c)
both types of dissipation are added with $\tau =\tau_\phi\approx 5 
\cdot 10^{-10}$s. When, as in d),  the frequency of the applied 
field is increased such that it 
approaches the superconductor energy gap, $2\Delta$,  we find that the system 
shows a {\it negative conductance}, see text, when $\tau =\tau_\phi\approx
2 \cdot 10^{-10}$s. For plots a,b,c $\hbar\omega/2\Delta\approx 0.75$ and 
for plot d $\hbar\omega/2\Delta\approx 0.95$. The amplitude of the external 
field is such that $|V_{+-}|/\Delta=0.024$ in all plots, and
for reference the case when $\tau\rightarrow\infty$ and $\tau_\phi\rightarrow\infty$ are included (dotted curves).\label{diss_plot}}
%\end{center}
\end{figure}

The first important property of this expression is that for 
$\tau_\phi\rightarrow\infty$ and $\tau \rightarrow\infty$ it reduces 
to the previous current expression that we derived, Eq.~(\ref{I}). One 
obvious effect on the current is that the oscillations of the current 
decrease when the dephasing time, $\tau_\phi$, becomes small. Another 
is that when the relaxation time, $\tau $, is decreased, the current 
diminishes, see Fig.~\ref{diss_plot}.

We note that the number of parameters has become quite large. It 
is now possible to calculate the current for different combinations of 
$\tau , \tau_\phi, V, V_{+-}, \omega,$ and the transmission coefficient 
$D$.  

A curious effect occurs when the relaxation time is such that the upper 
level, after being populated at point A, is fully depopulated at $\phi=\pi$, 
the middle of the Josephson period. 
What we find is a
%The result is a 
negative dc current for a positive bias, a regime where our system 
%could under special circumstances 
shows a {\em negative conductance}, see Fig.~\ref{diss_plot}, plots a and d. 
The relaxation time is such that the upper level is mainly populated when 
it's derivative is negative (it carries a negative current) and then the 
lower level is highly populated for the part of the period when it's 
derivative is negative. This effect will be most pronounced when the energy 
of the applied field, $\hbar \omega$, approaches the energy of the gap, 
$2\Delta$, as in Fig.~\ref{diss_plot}d. 

%The presence of a current which changes direction has been discussed 
%earlier, in a slightly  different situation~\cite{somer}. The method of 
%producing the current may be different, but the underlying physical mechanism 
%should be the same. 

Actually, the physical mechanism which is behind the appearance of this 
negative resistance is similar to the one responsible for the ``somersault
effect" effect discussed in Ref.~\cite{smslt}. In that case the method for
periodically bringing a pair of Andreev levels in
 a superconducting quantum point contact in resonance
with a microwave field was different. Rather than applying a voltage bias,
the time-dependence of the gate potential was assumed to have two components;
one slow component which shifts the Andreev levels through its effect on
the transparency $D=D(k_F(t))$ and one fast (microwave) component for coupling
the two levels.

%\begin{figure}[h]
%  \begin{center}
%    \centerline{\psfig{figure=Figures/neg_curr.eps,width=8cm}}
%    \caption{The current when $\omega=.9 \;2 \Delta$ and with a relaxation 
%time of $\tau =1.3\; 10^{-10}$s. There is a large negative current for 
%these values. In plot d) the frequency of the applied field is higher and we 
%have a 
%pronounced {\it negatice conductance}, see text and $\tau =\tau\phi\approx
%2 10^{-10}s.}
%\end{figure}
%      \end{center}
%\end{figure}
%\begin{figure}[h]
%  \begin{center}
%    \centerline{\psfig{figure=Figures/neg_curr2.eps,width=8cm}}
%    \caption{The current when $\omega=.82 \;2 \Delta$ and with a 
%relaxation time of $\tau =1.0\; 10^{-10}s$. In this more realistic 
%case for the frequency of the applied field a regime of negative 
%conductance can still be found.}
%      \end{center}
%\end{figure}
  
\section{Conclusions}
We have shown that irradiation of a voltage-biased
superconducting quantum point contact at frequencies $\omega\sim\Delta$ can 
remove the suppression of subgap dc transport through Andreev levels. 
Quantum interference among resonant scattering events can be used
for microwave spectroscopy of the Andreev levels. 

The same interference effect can also be applied for detecting
weak electromagnetic signals up to the gap frequency. Due to the resonant
character of the phenomenon, the current response is proportional to the
ratio between the amplitude of the applied field and the applied voltage,
$I\sim|V_\pm|^2/\Delta eV$. At the same time,
for common SIS detectors a non-resonant current response 
is proportional to the ratio between the amplitude and the frequency
of the applied radiation \cite{Tucker}), $I\sim |V_{\pm}/\omega|^2$, i.e. 
it depends entirely on the parameters of the external signal
and cannot be improved.

Finally, we note that
the classic double-slit interference experiment, where two spatially separated
trajectories combine to form an interference 
pattern, clearly demonstrates the wave-like nature of electron propagation.
For a 0-dimensional
system, with no spatial structure, we have shown that a completely 
analogous interference phenomenon
may occur between two distinct trajectories in the  {\em temporal} 
evolution of a quantum system. 
%
%Such
%trajectories may appear in the presence of temporally 
%localized non-adiabatic perturbations (rather 
%than spatially localized slits in a screen) which scatter the system from
%one adiabatically evolving state to another.
 
{\bf Acknowledgment.}
Support from the Swedish KVA, SSF, Materials consortia 9 \&
11, NFR and from the National Science Foundation under Grant No. 
PHY94-07194 is gratefully acknowledged. NIL and MJ are grateful for
the hospitality of the Institute for Theoretical Physics, UC Santa Barbara,
where part of this work was done.
%\vspace{-5mm}


\begin{thebibliography}{999}
\bibitem{LV}
This conclusion is in agreement with the analysis of multiple Andreev
reflections in Ref.~\cite{Lv1}.
\bibitem{Ave}
This result is obvious for a pure SNS contact ($D=1$) since the two
Andreev states involved in the transition at the band edge have momenta
$k_F$ and $-k_F$ pointing in opposite directions. See also~\cite{Ave1}.
\bibitem{C}
Shunting by the junction capacitance $C$ is small at relevant frequences
($\sim\!1/\delta t\!\ll\!\Delta$) if
$C\!\leq1\!/R_N\!\Delta\!\sim\!10^{-15}\!F$.
\end{thebibliography}
\end{document}